\documentclass[reprint,amsmath,amssymb,aps,prl,superscriptaddress]{revtex4-2}
\usepackage{graphicx} % Required for inserting images
\usepackage{chemformula}
\usepackage{siunitx}
\usepackage[utf8]{inputenc} 

\newcommand{\dtuphys}{Department of Physics}
\newcommand{\dtuaddr}{Technical University of Denmark, DK-2800 Kongens Lyngby, Denmark}

\begin{document}

%\title{Phonon edge states and their role in modelling High-Resolution Transmission Electron Microscopy data}
\title{Visualizing phonon edge states on molybdenum disulphide}
\author{Patrick Giese}
\affiliation{\dtuphys, \dtuaddr}
\author{Mathias Stokkebye Nissen}
\affiliation{Department of Energy Conversion and Storage, \dtuaddr}
\author{Stig Helveg}
\affiliation{Center for Visualizing Catalytic Processes (VISION), \dtuphys, \dtuaddr}
\author{Jakob Schiøtz}
\affiliation{\dtuphys, \dtuaddr}
\email{schiotz@fysik.dtu.dk}
\date{\today}

\begin{abstract}
We employ Molecular Dynamics (MD) simulations to study atom vibrational amplitudes in carbon-supported Molybdenum Disulphide (\ch{MoS2}) nanoparticles. Enhanced and correlated atom vibrational amplitudes are observed as the nanoparticle edges are approached from the bulk, consistent with recent experimental High-Resolution Transmission Electron Microscopy (HR-TEM) observations by Chen et al (Nature Communications 12, 5007 (2021). Analysis of phonon modes in finite systems explains the experimental observation by low-energy phonon modes confined at the nanoparticle edge, underscoring the need of full MD modeling for accurate HR-TEM image interpretation. Noticeably, we introduce a workflow for training Equivariant Neural Network-based machine learning potentials using limited Density Functional Theory (DFT) calculations. This approach effectively captures both covalent and van der Waals interactions, enabling accurate extrapolations of DFT calculations to larger systems with built-in error estimation.
\end{abstract}
\maketitle

%\section{Introduction}

Electron microscopy has become a powerful tool for visualizing matter at the atomic-scale.
Ultimately, it now reaches a spatial resolution down to ca.  50 pm in both broad- and focused-beam modes \cite{Kisielowski2008DetectionLimit,Erni2009Atomic-resolutionProbe}. This resolution level is dictated by the atomic potentials and their vibrational smearing rather than the electron optical performance \cite{Chen2021Probing3D,Chen2021ElectronVibrations}. However, while atom vibrations are commonly described using an independent atom model \cite{Loane1991ThermalDiffraction,Hillyard1995DetectorImaging}, Chen \emph{et al.} \cite{Chen2021Probing3D} related a measured phase contrast decay from the bulk toward the edge of a single-layer \ch{MoS2} nanoparticle on a few-layer graphene support to vibrations that must be correlated. 
 
Here we examine the role of atomic vibrations on the high-resolution transmission electron microscopy (HRTEM) images using molecular dynamics (MD) simulations.  Specifically, we follow the frozen-phonon approach in which atoms are displaced around their equilibrium sites in the sample and images of a large number of such configurations are generated and averaged to form the actual image.  However, rather than drawing random displacements for each atom from a Gaussian distribution, corresponding to independently vibrating atoms in an Einstein model, we use MD simulations with a machine learning potential to describe realistic and correlated atomic vibrations.

There are two ways to include vibrations in simulations of HR-(S)TEM images.  One can replace the electrostatic potential of the sample with a time-averaged potential for example by smearing the potentials of the individual atoms with a Debye-Waller factor.  Alternatively, one can randomly displace atoms in the sample and generate a large number of images of such configurations before averaging the resulting images.  This so-called \emph{frozen phonon method} \cite{Loane1991ThermalDiffraction,Hillyard1995DetectorImaging,VanDyck2011PersistentMicroscopy} correctly captures the separate time scales of the imaging process.  Each electron passes through the sample with a speed close to the speed of light, leading to an interaction time much shorter than the vibrational time of the atoms.  On the other hand, the time between two electrons passing through the sample is much larger than the vibrational time, each electron thus images a different snapshot of the atomic configurations \cite{Loane1991ThermalDiffraction}.  Applying a Debye-Waller factor corresponds to incorrectly assuming that the interaction time of the electrons is long compared to the vibrational time.

In the usual implementations of the frozen phonon method each atom is displaced independently; this corresponds to the Einstein model for vibrations in solids, where atoms are uncoupled harmonic oscillators.  In reality vibrations are collective waves with vibrational amplitudes depending on the local structure being bulk, surface, interface or other defective sites.

In this Letter, we explore the vibrational modes in \ch{MoS2} nanoparticles in the context of the observations of Chen \emph{et al.}  We present a systematic method for training a machine learning potential based on active learning \cite{Smith2018LessLearning} and Equivariant Neural Networks \cite{Thomas2018TensorClouds,Batzner2022E3-equivariantPotentials} for a supported nanoparticle, in this case graphite supported \ch{MoS2}, including a validity assessment of the potential in subsequent simulations.  We then use Molecular Dynamics (MD) to investigate the vibrational amplitudes of the atoms in the nanoparticle.  In agreement with the observations by Chen \emph{et al.}, we find an increased vibrational amplitude not just for the under-coordinated edge atoms, but also for atoms a few lattice constants away from the edge.  We use phonon simulations to show that this is due to the existence of edge states where the majority of the vibrational amplitude is localized near the edges.

\begin{figure}
    % Made by notebook PatrickPlots.ipynb in folder 
    % /home/niflheim/schiotz/simulations/MoS2_phonons_new
    \centering
    \includegraphics[width=\linewidth]{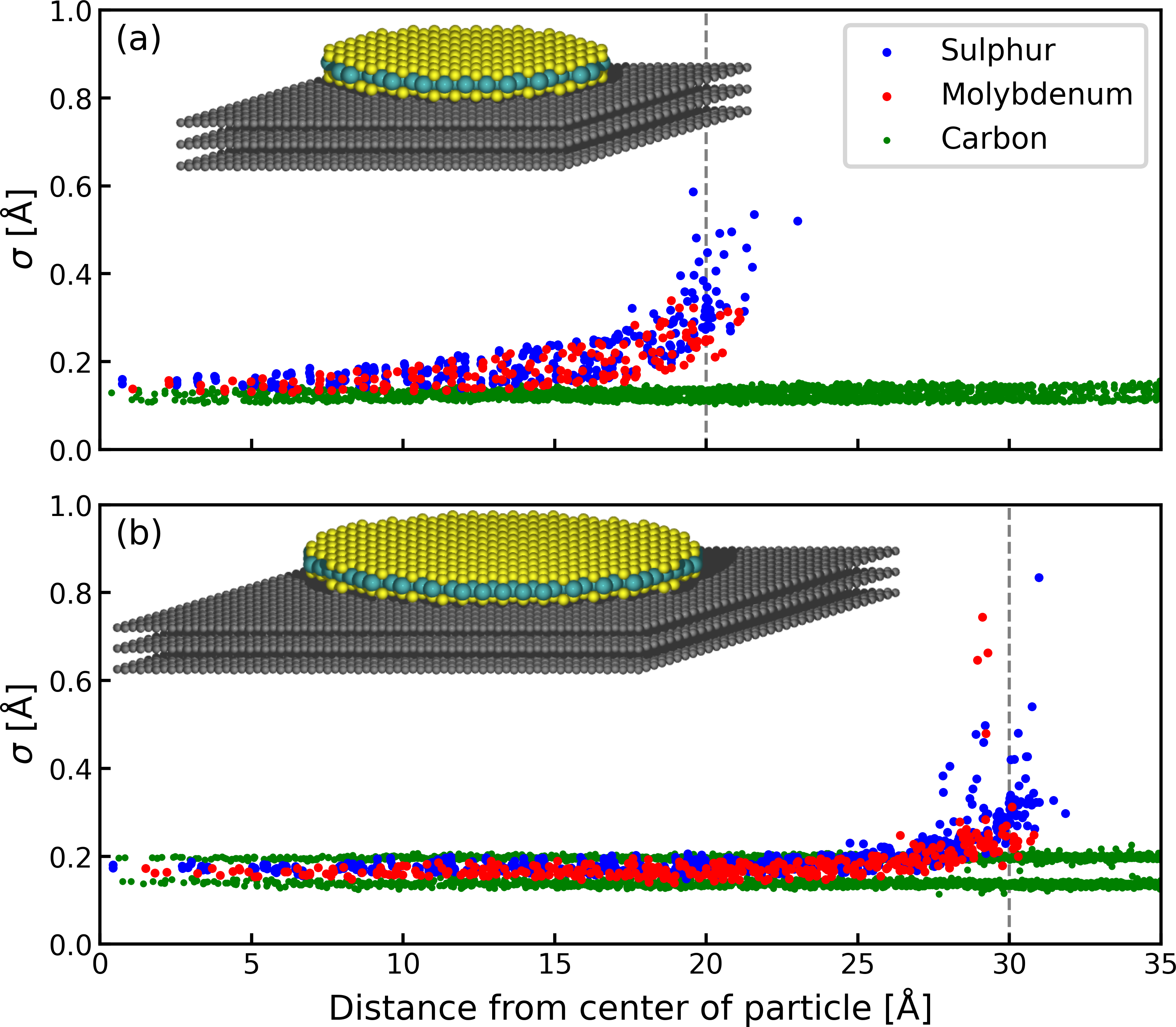}
    \caption{In-plane atomic vibrational amplitudes across \ch{MoS2} nanoparticles and  supporting bi-layer graphene. The graphs display the vibrational amplitude of Sulphur (blue), Molybdenum (red), and Carbon (green) atoms as a function of distance from the center of the nanoparticles, for a nanoparticles with radius 2 nm (panel a) and 3 nm (panel b).  The initial configurations of the nanoparticles are shown as insets. The temperature is $\SI{300}{\kelvin}$.  The vibrational amplitudes of Mo and S atoms increase close to the edge of the nanoparticles.  The carbon atom form two lines, as the lowest layer of the graphite was kept fixed, and the atoms in the layer just above the fixed atoms show lower vibrational amplitude in the transverse direction.}
    \label{fig:vibrationamplitudesA}
\end{figure}

We optimized equivariant neural network potentials (ENNPs) describing molybdenum disulphide (\ch{MoS2}) supported by graphite or graphene, as described below.  The potential was subsequently used for Molecular Dynamics (MD) simulations of graphite-supported \ch{MoS2} nanoparticles.  Circular nanoparticles with diameter \SI{30}{\angstrom} and \SI{20}{\angstrom} were created, as shown in Fig.~\ref{fig:vibrationamplitudesA}.  In addition, nanoparticles of the same size but with more sulphur on the edge were created by using a marginally larger cutoff radius for sulphur than for molybdenum.

Molecular dynamics simulations were run in the canonical (NVT) ensemble at temperatures of 300-500 K.  The first 5 ps were thermalization and stabilization, the following 5 ps were used to collect vibrational amplitudes, defined as the standard deviation of the atomic positions, $\sigma_n = \sqrt{\left<\mathbf{r_n}^2\right> - \left<\mathbf{r_n}\right>^2}$, where $\mathbf{r_n}$ is the position of the $n$'th atom, and the means are taken over the time series.  

In Fig.~\ref{fig:vibrationamplitudesA}, the vibrational amplitudes are plotted versus radial position in the nanoparticle, for the different atomic species.  We clearly see an increase in vibrational amplitude for both Mo and S atoms located near the edge of the nanoparticle.  The vibrational amplitudes of the outermost sulphur atoms increase slightly when the edge is sulphur rich, see appendix Fig.~\ref{fig:vibrationamplitudesB}.

The increase in vibrational amplitudes near the edge of the nanoparticle is consistent, at least qualitatively, with the recently published data of Chen \emph{et al.}~\cite{Chen2021Probing3D}, where they find an increase in vibrational amplitudes near the edge of \ch{MoS2} nanoparticles.  They recorded a focal series of HRTEM images and reconstructed  the  wave function of the electron beam exiting a specimen consisting of single-layer \ch{MoS2} nanoparticles on few-layer graphene support \cite{Chen2021Probing3D}.

\begin{figure}
    \centering
    \includegraphics[width=\linewidth]{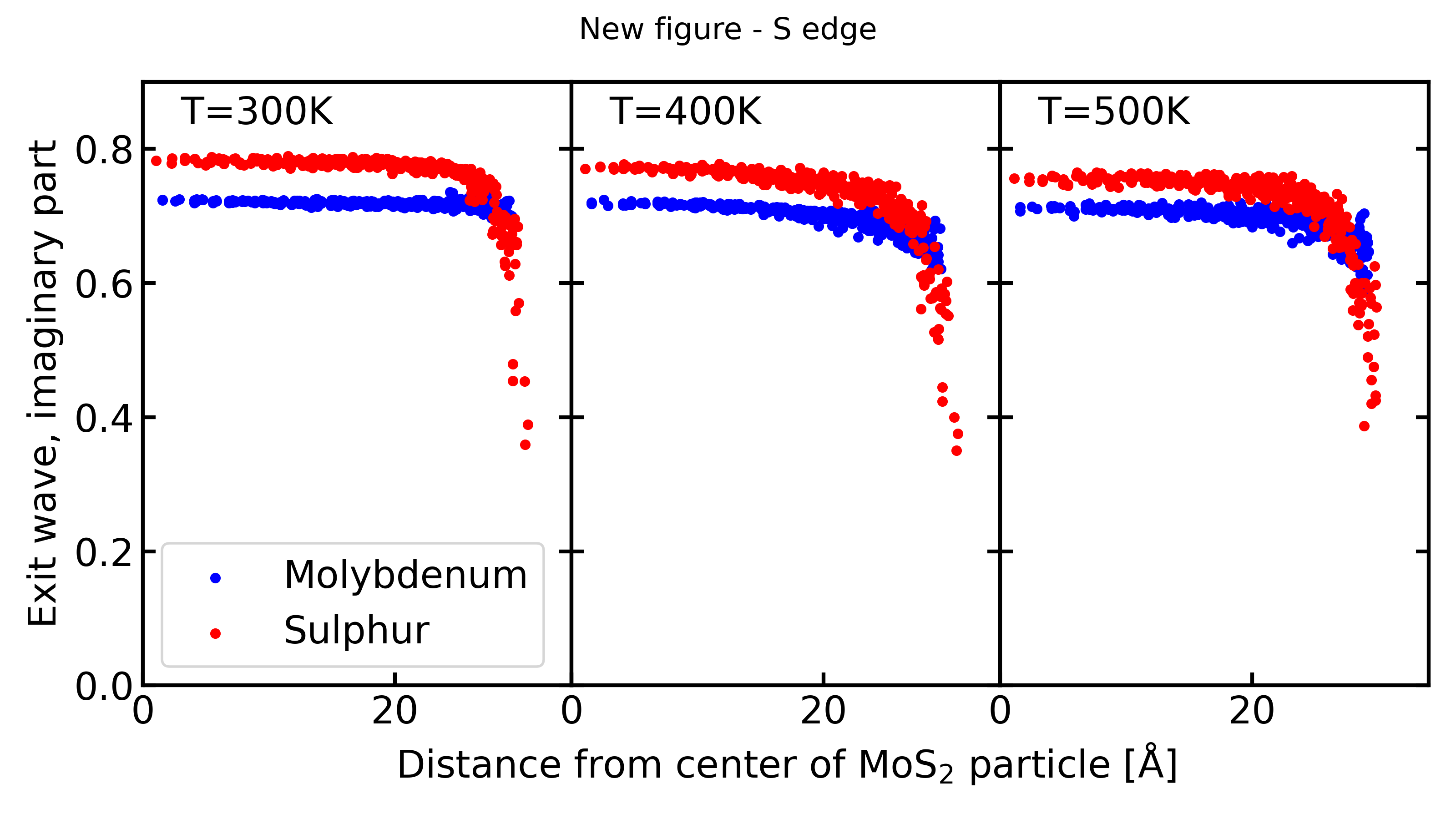}
    % Made by PlotExitwaveData.ipynb and MakeExitwaveData.ipynb
    \caption{The imaginary values of the peaks in the exit wave simulations for the S-rich sample at different temperatures.  The peak values are plotted for all peaks in the exit wave, as a function of the distance from the nanoparticle center.}
    \label{fig:imagvalues}
\end{figure}

\begin{figure*}
    % Figure made by notebook PhononStrip_Wide.ipynb in folder
    % /home/niflheim/schiotz/simulations/MoS2_phonons_new
    \centering
    \includegraphics[width=\linewidth]{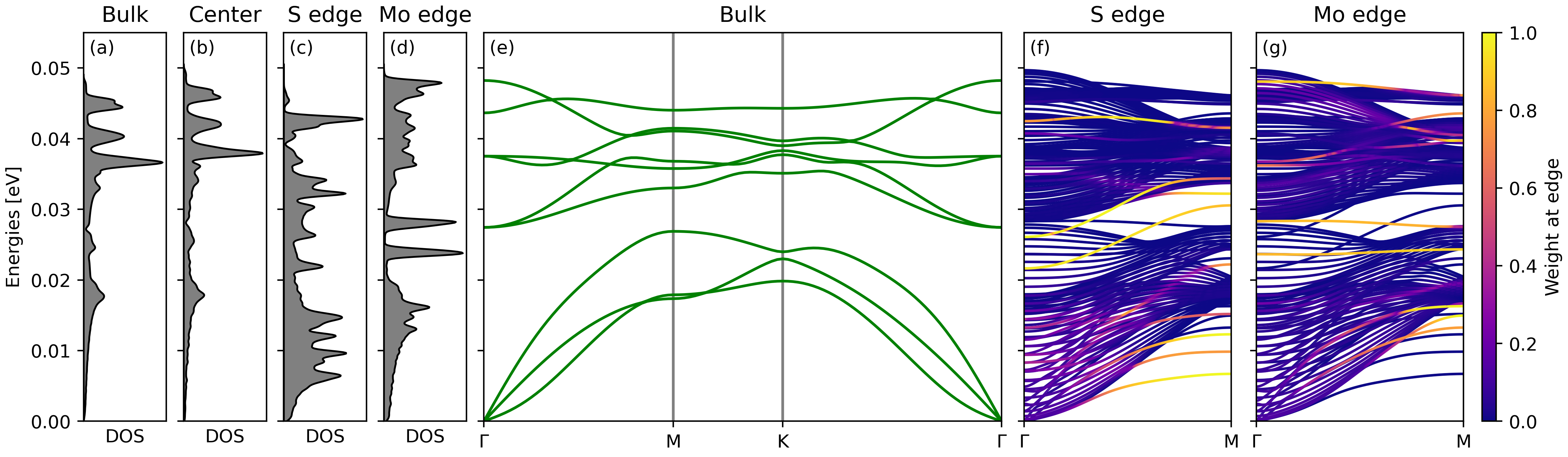}
    \caption{The phonon state (DOS) in \ch{MoS2} in bulk and near the edges.  Panel (a) shows the bulk phonon DOS.  Panel (b) shows the phonon DOS of a strip of \ch{MoS2} 20 unit cells wide, projected onto the central unit cell of the strip.  It is all but indistinguishable from the bulk DOS.  Panels (c) and (d) show the same DOS projected on the formula unit closest to the Mo and S edge, respectively.  In both cases we see a significant increase of the DOS at low energies.  Panel (e) shows the bulk band structure corresponding to the DOS in panel (a).  Panel (f) and (g) show the phonon states in the strip, colored by their weight (squared amplitudes, see text) on the sulphur and molybdenum edges, respectively.}
    \label{fig:PhononDOS}
\end{figure*}

In order to compare more directly with the results by Chen \emph{et al.}, we simulated the exit wave from our simulation, using the abTEM simulation software \cite{Madsen2021ThePrinciples} (see appendix).  We modelled atomic vibrations within the frozen phonon approximation, using 100 snapshots of the atomic positions from the last half of our MD simulation.  The electron exit wave imaginary part peaks at atomic column positions with a value that represents the projected electrostatic potential averaged over the area of the smeared atoms in the column \cite{Chen2021Probing3D}. Figure \ref{fig:imagvalues} shows the exit wave imaginary peak values from the center to the edge of the 3 nm wide \ch{MoS2} nanoparticle in Fig. \ref{fig:vibrationamplitudesA}(b).

While it might be expected that the peak values are lowered at the edge, because enhanced atom vibration and thus atom smearing results from the reduced atom coordination at the edge, it is surprising that the decrease in exit wave amplitude is seen up to a nanometer or more from the edge of the nanoparticle.  This finding reflects an enhanced vibrational amplitude both in experiment and in the MD simulations, and could indicate softer phonon modes near the edge of the sample.

To investigate this further, we have simulated phonons in \ch{MoS2} both in the infinite 2D sheet and in a strip of the material, using the same ML potential.  The results are summarized in Figure \ref{fig:PhononDOS}(a-d), where the phonon Density of State (DOS) is compared for an infinite 2D sheet of \ch{MoS2} and for a strip 20 formula units wide.  In the latter case, the DOS is projected onto the formula unit (one Mo and two S atoms) at the center of the simulation, and onto formula units at the two edges of the \ch{MoS2} strip.  While the projected DOS near the center of the strip is almost indistinguishable from the bulk DOS, a strong shift towards lower frequencies is seen at both edges.  Phonon states with energies below $k_BT \approx \SI{0.025}{\electronvolt}$ will contribute most strongly to the vibrational amplitudes, and this is where we see the increase in DOS.  By \emph{projected DOS} we mean the density of states where each state is weighted by the squares of the vibration amplitudes of the atoms we project onto.

The finite width of the strip removes the dispersion of the phonon modes perpendicular to the strip, and replaces it with discrete modes resulting in a larger number of bands, these are plotted in figure \ref{fig:PhononDOS}(f-g) colored by the weight of the modes on the two edges of the strip.  The weight is defined as the sum of the squared amplitudes on the three atoms (one Mo, two S) nearest the edge, divided by the sum over all atoms.  We clearly see that some bands are strongly localized at one of the edges of the strip, resulting in the increase of DOS observed at the edges.  The DOS is affected to a lesser degree up to two unit cells from the edge, see appendix (Fig.~\ref{fig:compareDOS}).

\begin{figure}
    % Made by notebook PlotPhononModes.ipynb in folder
    % /home/niflheim/schiotz/simulations/MoS2_phonons_new
    \centering
    \includegraphics[width=\linewidth]{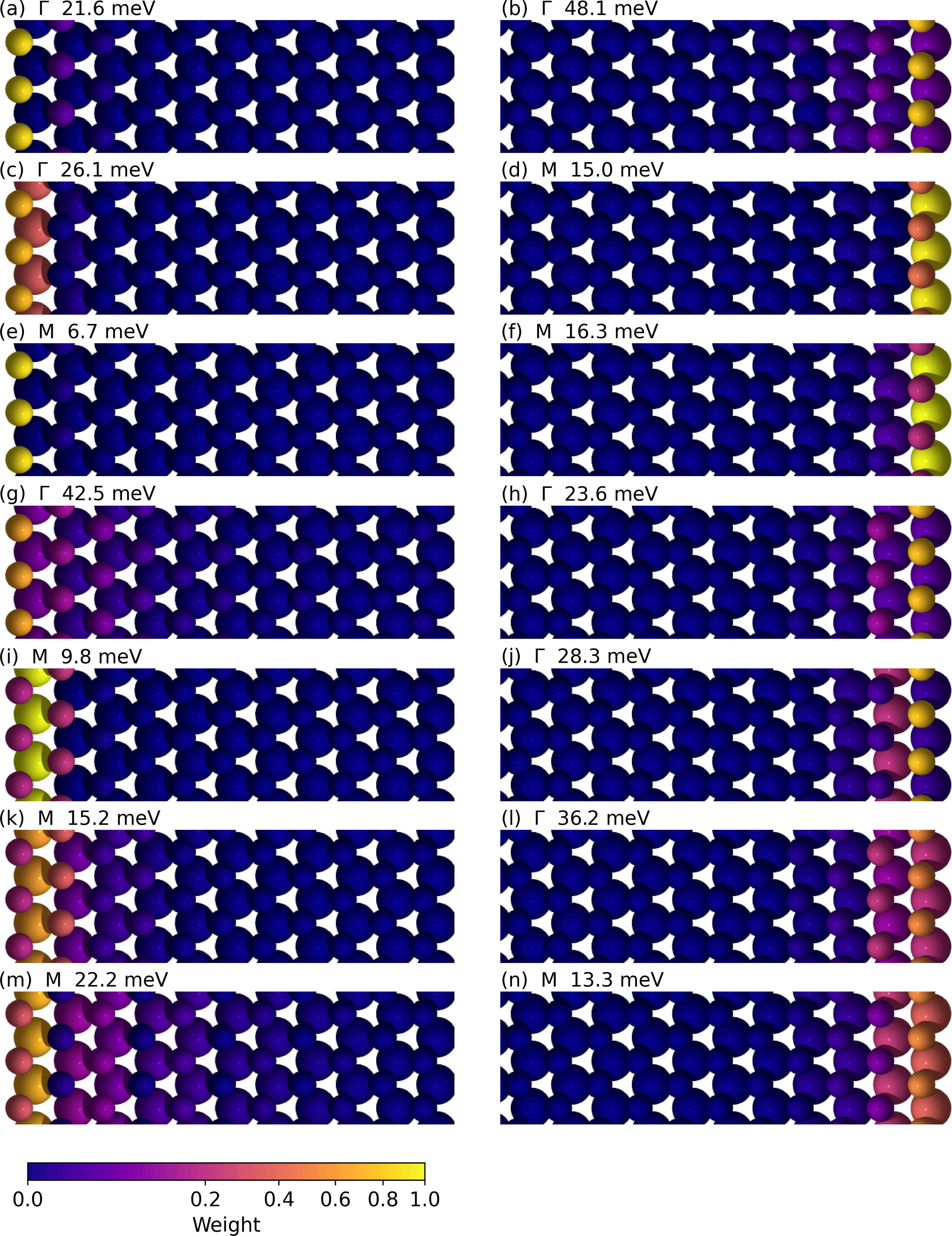}
    \caption{Phonon modes localized at either the sulphur edge (left) or the molybdenum edge (right).  Panel (a)-(f) show the most localized modes.  Panel (g)-(n) show modes that extend a few lattice constants away from the edge.  Only modes at the center of the Brillouin zone (marked $\Gamma$) or at the edge (marked M) are shown.  The color shows the norm-square weight of the mode at each atom.}
    \label{fig:phononmodes}
\end{figure}

Selected modes at the $\Gamma$ and X points are shown in Figure \ref{fig:phononmodes}.  The top three modes at each edge are examples of modes that are strongly localized at the edge. The lower four modes illustrate that there exist modes that extend a few lattice constants away from the edge.  We propose that these modes are responsible for the increase in vibrational amplitude observed both experimentally and in our simulations.  Animations of the modes in Fig.~\ref{fig:phononmodes} can be found in the Supplementary Material \cite{Supplementary}.

%\clearpage
%\section{Method}

%\subsection{Generation of the machine learning potentials}
\emph{Generation of the machine learning potentials} --- 
The simulations are based on Equivariant Neural Network Potentials (ENNPs) as implemented in the NequIP package \cite{Batzner2022E3-equivariantPotentials}.  NequIP is an equivariant graph neural network, where the energy of each atom is calculated from the vectors to the neighboring atoms and their chemical species.  Unlike invariant neural network potential, which requires $10^4$ -- $10^7$ structures in the training set \cite{Schutt2017SchNet:Interactions,Smith2017ANI-1:Cost,Mikkelsen2021IsTemperature}, ENNPs can typically be trained on a few hundred training structures \cite{Batzner2022E3-equivariantPotentials,NunezValencia2024BeamPotentials}.

We use a hierarchical training procedure, where each generation of the neural network is used to identify relevant structures to add to the training set of the next generation.  The initial generation 0 is created from four nanoparticles of \ch{MoS2} containing from 9 to 24 atoms, two terminated by sulphur atoms and two by molybdenum atoms.  From these 100 different structures were generated, by placing them on 1--3 graphene layers and randomly removing some atoms, see appendix for details.

The 100 generated structures are then energy minimized with Density Functional Theory (DFT).  We use the GPAW \cite{Enkovaara2010ElectronicMethod,Mortensen2024GPAW:Calculations} electronic structure code, the PBE exchange correlation potential \cite{Perdew1996GeneralizedSimple} with D3 dispersion \cite{Grimme2010AH-Pu}, with parameters described in the appendix.  The energy of the structures are then minimized with the BFGS algorithm as implemented in the Atomic Simulation Environment (ASE) \cite{HjorthLarsen2017TheAtoms} until the force is below \SI{0.2}{\electronvolt\per\angstrom}, resulting in a final configuration with small but non-zero forces for the training of the ENNP.  The first, the middle and the last configuration of each trajectory are used for training.  Of the 300 configurations thus generated, the last 50 are saved for the test set.  The  remaining 250 are the training set, which is then split randomly into five parts, and five different ENNPs are trained by leaving out one fifth of the data for each.  These potentials are \emph{generation 0}.

Each successive generation of potentials is then made by generating 10--18 new initial structures.  Each is used as an initial configuration for a short MD simulations at constant temperature, using the average energy and forces of the five potentials trained in the previous generation \cite{Smith2018LessLearning}.  For each MD trajectory, we selected three configurations were the five potentials disagreed the most on the forces, discarding situations where the forces were unreasonably large, as described in the appendix.

We repeated this procedure until no further improvement in the potential was seen, this occurred after \emph{generation 5}; with \emph{generation 6} we saw a slight increase in the test error.

An important lesson learned while developing this scheme was the importance of removing configurations with unusually large forces (above \SI{100}{\electronvolt\per\angstrom} as described in the appendix). If such forces were not removed, the error of the forces predicted by the resulting potential would be more than an order of magnitude larger, even on a test set that does include such large-force configurations.

A useful modification to the workflow described above could be to replace the DFT energy minimizations in \emph{generation 0} with either energy minimizations or short molecular dynamics simulations using one of the universal machine learning potentials recently published \cite{Deng2023CHGNetModelling,Batatia2024AChemistry}.  In this case three selected configurations will still be picked from each starting system, but a DFT calculation will only have to be done for the selected configurations, not for the entire trajectory.  It is of course important not to train the new machine learning model on energies and forces calculated by the approximate model, but on forces and energies calculated by DFT in a configuration generated by the approximate model.

The final potential used for the simulations was trained on the entire \emph{generation 5} training set.  In addition, the five different potentials trained on 4/5th of the \emph{generation 5} data set were kept for \emph{a posteriori} error estimation.

%\subsection{Nanoparticle simulations}
\emph{Nanoparticle simulations} ---
Two circular nanoparticles of \ch{MoS2} were created by cutting out atoms from a flat sheet, with cut-out radii of \SI{20}{\angstrom} and \SI{30}{\angstrom}.  As the nanoparticles thus created had rather sulphur-deficient edges, an additional two nanoparticles were made with the same cut-off radius used for the sulphur atoms, but the cutoff radius radius was reduced by \SI{1}{\angstrom} for the molybdenum atoms.

The resulting four nanoparticles were placed on three-layer slabs of graphite, and MD was performed at temperatures of \SI{300}{\kelvin} -- \SI{500}{\kelvin}, using Langevin dynamics.  In all cases, the positions of the atoms in the lowest graphene layer were kept fixed.  As simulating using an ensemble of five potentials slows the simulation by a factor five, we used the single potential trained on the entire \emph{generation 5} data set, and only used the ensemble for validation of shorter sequences.

\begin{figure}
    %\centering
    \includegraphics[width=\linewidth]{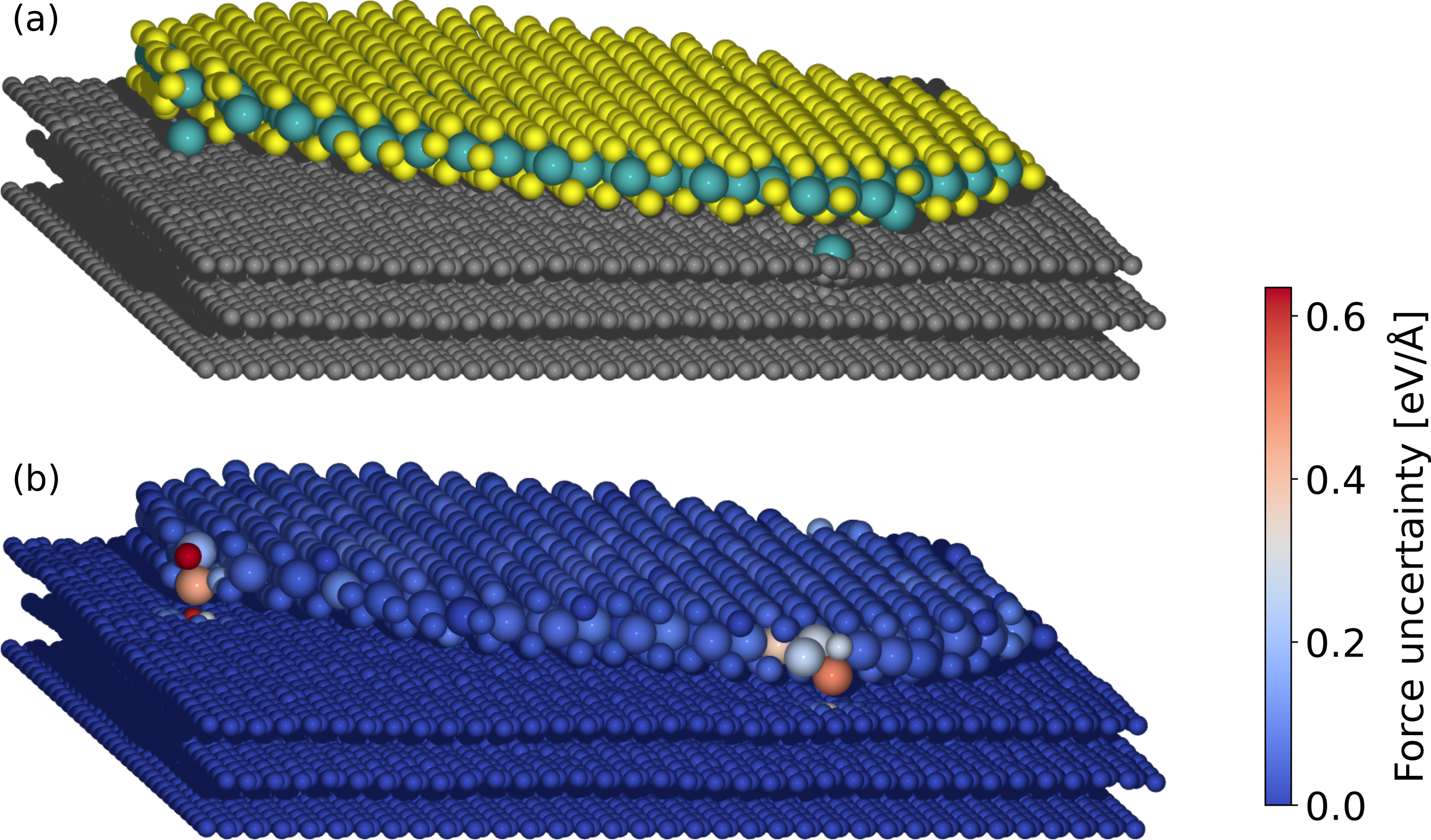}
    % Generated by notebook PatrickPlots.ipynb
    \caption{(a) An example of an error produce by the model potential: a few Mo atoms migrate onto the graphite substrate.  (b) The variation of the force between five models just as the weird configuration is beginning to appear, clearly demonstrating that the models are outside their training data and essentially guessing.  Cutting out the regions around the problematic atoms and adding the two cut-out regions to the training set eliminates the problem.}
    \label{fig:th29}
\end{figure}

During the simulations, we observed a single case of a chemically unreasonable reaction taking place, where two molybdenum atoms left one of the nanoparticles and migrated onto the graphite substrate, see Fig.~\ref{fig:th29}.  In this case we could confirm that the system had passed through a configuration where five ENNPs disagreed on the force on the affected atoms, confirming that we had encountered a configuration outside the confidence regime of the potential.

We cut out regions around the problematic atoms, by including atoms in cylinders with radii \SI{7}{\angstrom}, \SI{10}{\angstrom} and \SI{15}{\angstrom} centered on the two problematic Mo atoms.  These six configurations were added to the training set, and a new potential was fitted.  Re-running the simulations with the new potential confirmed that no problematic configurations were encountered.

%\section{Conclusion}
\emph{In conclusion, }
we have presented a workflow for training machine learning potentials for specific simulation problems, and have trained a potential for carbon supported molybdenum disulphide nanoparticles.  The potential is based on equivariant graph neural networks as implemented in the Nequip package.  The targeted training workflow makes it possible to get well-performing networks with only slightly more than 400 DFT-calculated configurations in the training set.  By using an ensemble of five neural network potentials (``query by committee''), the quality of the potential can be monitored even for system sizes where DFT calculations are prohibitively expensive.

The potential is used to investigate thermal vibrations in supported nanoparticles, and show that the vibrational amplitudes are significantly enhanced in a region near the edge of the nanoparticle.  This is in agreement with recent observations with recent HR-TEM observations, and can be explained by localized phonon modes near the edges of the two-dimensional nanoparticles.  

This illustrates the importance of using molecular dynamics simulations to correctly model the vibrations when modelling HR-TEM images, and we provide a procedure for obtaining the machine learning potentials required to perform such simulations.

\emph{Acknowledgments} ---
The authors acknowledge financial support from the Independent Research Fund Denmark (DFF-FTP) through grant no. 9041-00161B.  The Center for Visualizing Catalytic Processes (VISION) is funded by the Danish National Research Foundation (DNRF146).   We acknowledge support from the Novo Nordisk Foundation Data Science Research Infrastructure 2022 Grant:  A high-performance computing infrastructure for data-driven research on sustainable energy materials, Grant no. NNF22OC0078009.

%\bibliographystyle{unsrtnat}
% cleanedreferences.bib is produced from references.bib by the
% script cleanrefs.py.  It is most easily done by syncing with GIT,
% and running it offline.

%\bibliography{cleanedreferences}

\appendix

\section{Appendix: computational details}

\begin{figure}
    \centering
    \includegraphics[width=\linewidth]{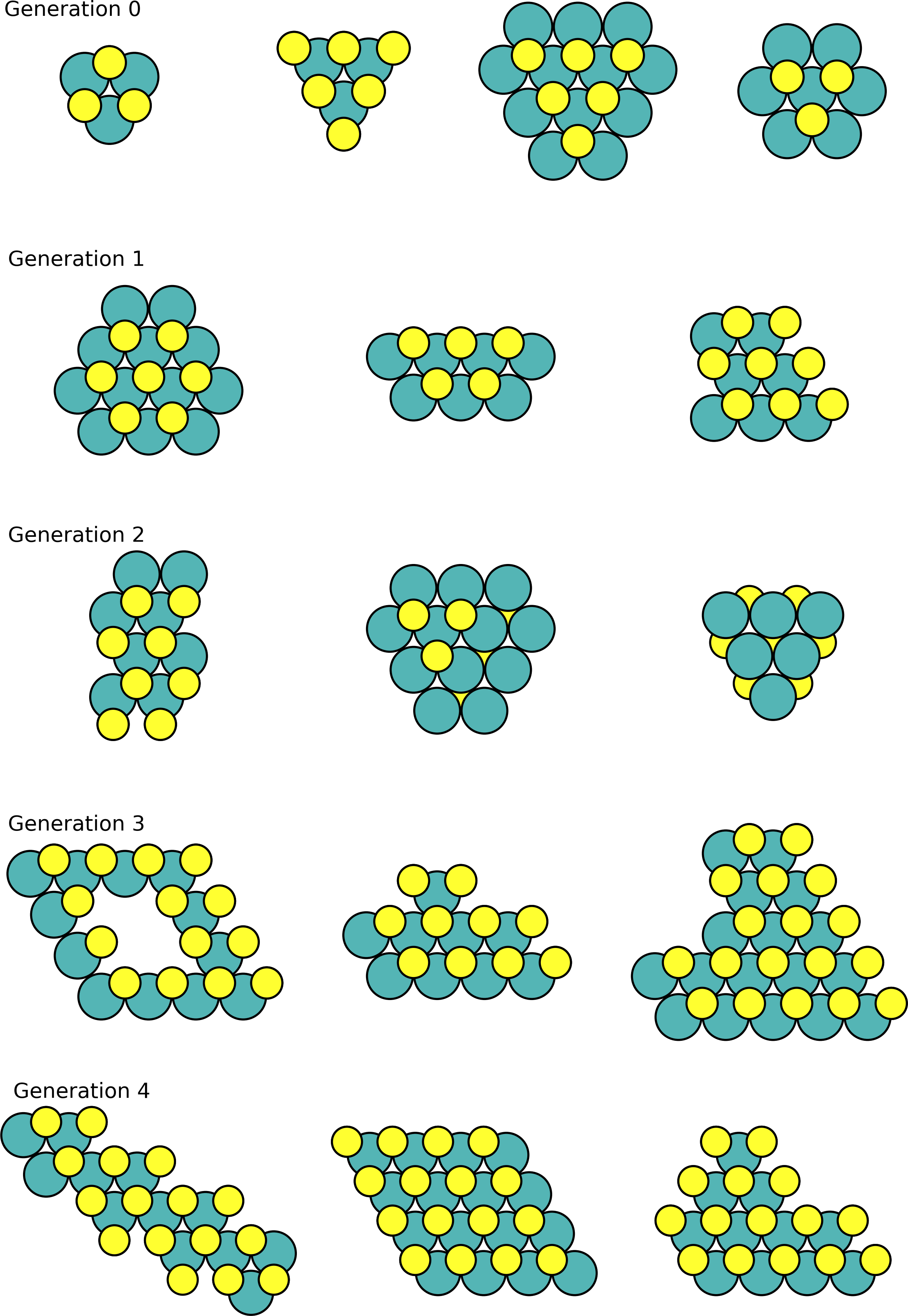}
    \caption{Initial \ch{MoS2} clusters used for generation 0--4.  Generation 5 and 6 reused the clusters from generation 0 and 1.  Some of them are physically unrealistic, but have value to generate a diverse training set.}
    \label{fig:initclusters}
\end{figure}

%\subsection{Generation of structures for machine learning}
\emph{Generation of structures for machine learning} --- 
For each generation, three or four clusters of \ch{MoS2} are generated, see Fig.~\ref{fig:initclusters}.  From 10 to 100 supported systems are then generated (100, 10, 18, 18, 18, 18 for generations 0--6)  by randomly picking one of the  clusters and placing it on one to three layers of graphene.  The distances between graphene layers as well as the distance between the nanoparticle and the graphene is varied slightly around the van der Waals minimum distances.  The system is periodic in the $xy$ plane, so the support forms an infinite sheet.  The supercell is chosen large enough that the nanoparticles do not interact with themselves through the periodic boundary conditions.  All atoms are then displaced by a random amount from a Gaussian distribution with spread \SI{0.1}{\angstrom}.  Finally, in one third of the samples two atoms are randomly removed from the \ch{MoS2}, in one third five random carbon atoms are removed, and in the remaining third all atoms are kept.  

\emph{Density Functional Theory calculations} ---
We use the GPAW electronic structure code \cite{Enkovaara2010ElectronicMethod,Mortensen2024GPAW:Calculations}, and the PBE exchange correlation potential \cite{Perdew1996GeneralizedSimple} with D3 dispersion \cite{Grimme2010AH-Pu} added to properly describe the van der Waals interactions between the layers.  The Kohn-Sham wavefunctions were described in a plane wave basis with a cutoff of 600 eV.  As the supercell is large in the periodic directions, the Brillouin zone is sampled with a single k-point, the $\Gamma$ point.  As the forces on the atoms converge more slowly in the Kohn-Sham scheme than the energies and the electron density, and as the forces are explicitly needed to fit the machine learning potential, an additional convergence criterion was added to the Kohn-Sham self consistency loop, requiring force convergence within \SI{0.01}{\electronvolt\per\angstrom}.  

In generation 0, the energy of the structures are minimized with the BFGS algorithm as implemented in the Atomic Simulation Environment (ASE) \cite{HjorthLarsen2017TheAtoms} until the force is below \SI{0.2}{\electronvolt\per\angstrom}.  This very loose convergence criterion ensures that the energy minimization ends with a structure close to but not at a local energy minimum, such that there are non-zero forces available for the training of the neural network potential.  

In generation 1--6, no energy minimization is done as the configurations are realistic MD configurations.  The forces and energies are just calculated and used for training.

\begin{figure*}[t]
    % This figure is placed early to make it appear at the top of a page
    \centering
    \includegraphics[width=\linewidth]{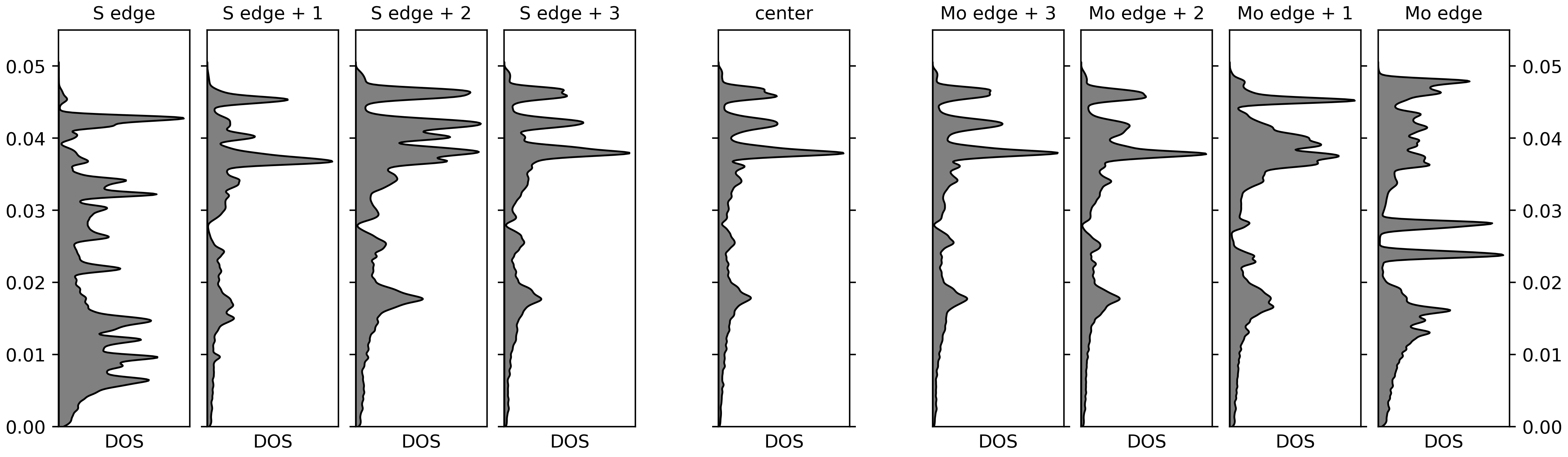}
    % Generated by notebook PhononStrip_Wide.ipynb
    \caption{The phonon state (DOS) in \ch{MoS2} is perturbed by the edge state.  The perturbation is clearest in the unit cells at the edge, but can be seen up to two unit cells away from it.}
    \label{fig:compareDOS}
\end{figure*}

\emph{Molecular Dynamics} ---
MD was used in the same way when generating structure in the active learning procedure, and for the final simulations.  The Langevin algorithm as implemented in ASE \cite{HjorthLarsen2017TheAtoms} is used for temperature control during the molecular dynamics, the time step is set to \SI{1}{\femto\second} and the Langevin friction was \SI{0.02}{\sqrt{\electronvolt\per u}\per \angstrom} corresponding to a thermalization time constant of \SI{0.51}{\pico\second}.

\emph{Active Learning} ---
MD was run for 400 time steps using the average of the five potentials trained in the previous generation \cite{Smith2018LessLearning}. The temperature is chosen randomly to either \SI{300}{\kelvin}, \SI{400}{\kelvin} (for generation 1-3, only \SI{400}{\kelvin} was used).  
For each step in the MD simulation, we calculated the standard deviation of the forces predicted by the ensemble of models.  This produced a standard deviation for each atom, the average of these were then used to detect configurations where the five models disagreed.

As the early generations of the potential were not reliable, occasionally absurdly large forces would be predicted, often due to the appearance of physically unreasonable structures.  If forces above \SI{100}{\electronvolt\per\angstrom} appeared in an MD simulation, the configurations from that point on would be discarded and not used in generating training data.

After this filtering, we selected the three configurations of each MD trajectory with the largest standard deviation between the five predicted, but with the additional constraint that no two configurations less than six time steps apart were selected.  These new configurations were added to the training set, and a new generation of five models was trained.

\emph{Exit wave simulations} ---
To compare with the experimental exit waves \cite{Chen2021Probing3D}, we simulated thermally averaged exit waves from the MD simulations, using the multislice algorithm and the frozen phonon method.

The frozen phonon method leverages the separation of time scales in electron microscopy.  The time scale of the camera (\si{\micro\second} to \si{\milli\second}) is very long compared to the atomic vibrations (sub-\si{\pico\second}), which again are slow compared to the time it takes for an electron to traverse the sample (tens of \si{\atto\second}).  The final image should thus be an average of individual images calculated for independent snapshots of the atomic vibrations.  This is in contrast with the method of applying a Debye-Waller factor, which corresponds to calculating a single image of an averaged atomic potential.

It should be noted that in the frozen phonon approximation, there is not a single exit wave of the sample, but an exit wave of each atomic configuration.  Nevertheless, we calculate an \emph{averaged} exit wave, for use in comparing to the exit wave reconstructed from the experimental image sequence.

The exit wave is calculated using the multislice algorithm as implemented in the abTEM software \cite{Madsen2021ThePrinciples}.  As we compare to reconstructed exit waves where the support has been removed by Fourier filtering the images prior to the exit wave reconstruction, we need to simulate an exit wave without the support.  We thus remove all carbon atoms from the simulations.  We select 100  distributed configurations from the second half of the MD simulation, for each of these the exit wave is calculated with abTEM, and are subsequently averaged.  The multislice simulation is performed with a lateral grid size of 0.05 Å, and a slice thickness of 0.5 Å.  The atomic potentials are parameterized according to Lobato and Van Dyck \cite{Lobato2014AnConstraints}.

Once the averaged exit wave is calculated, we find local peaks in the imaginary part of it.  The peak values and their distance from the center of the nanoparticles are then extracted, and used to produce Fig.~\ref{fig:imagvalues}.

\emph{Phonon band structure simulations} ---
The phonon band structures and densities of state were calculated using the \texttt{Phonons} module of the ASE, using the same ENNP used for the simulations.  The phonon spectrum is calculated from a single free-standing layer of \ch{MoS2}, this results in a transverse acoustic mode with quadratic dispersion, corresponding to buckling of the 2D sheet of material.  This may distort the very lowest parts of the density of states.  Unfortunately, calculating the phonon band structure of a supported material is not possible as the unit cell of the material and support do not match.

To investigate the existence of modes localized at the edges, we performed similar phonon calculations with a strip of \ch{MoS2} 20 unit cells wide.  This system thus only has periodic boundary conditions along one direction along which the phonons may propagate.  In the transverse direction, standing waves will form, splitting each bulk phonon band into 20 new bands.  The resulting band structure is therefore quite ``messy``, as seen in Fig.~\ref{fig:PhononDOS}.

The DOS projected on the formula units at the edge of the strip is significantly altered, as already shown in Fig.~\ref{fig:PhononDOS}.  As the edge phonon modes extend a few unit cells into the unit cell, so does the perturbation of the projected DOS.  This is seen in Fig.~\ref{fig:compareDOS}, where a perturbation can be seen up to two unit cells from the edge.

\begin{figure}
    \centering
    \includegraphics[width=\linewidth]{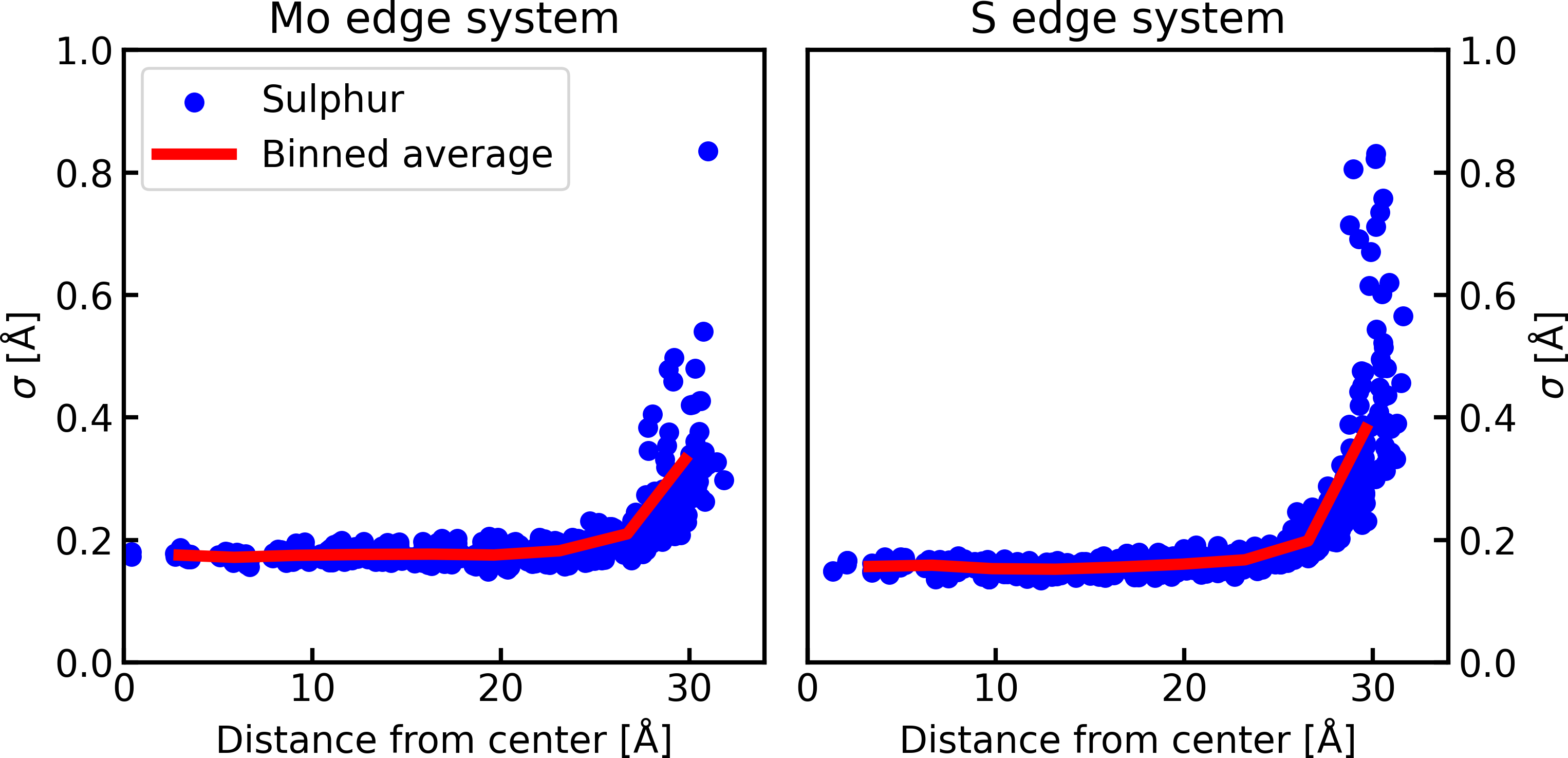}
    \caption{The vibrational amplitude of Sulphur atoms for nanoparticles with Mo-rich and S-rich edges.}
    \label{fig:vibrationamplitudesB}
\end{figure}

%\subsection{Difference between sulphur poor and sulphur rich nanoparticles.}
\emph{Effect of edge termination} ---
The simulations were made with nanoparticles where the edge was Mo-rich or S-rich.  The difference is minimal; the outermost sulfur atoms are more loosely bound in the S-rich particles, and thus have a higher vibrational amplitude.  That is shown in Fig.~\ref{fig:vibrationamplitudesB}.

%This is reflected in the exit wave, where Fig.~\ref{fig:imagvaluesMo} show the imaginary values of the exit wave for the Mo rich samples, to be compared directly with Fig.~\ref{fig:imagvalues} (S rich).  The difference is again minimal.

% \section{Supplementary Online Information: Animations of phonon modes.}
%
% Animated renderings of the phonon modes in Fig.~\ref{fig:phononmodes} can be found online at XXX.  The files are named by the edge, the k-point and the energy in meV.

\end{document}